# Analytical models of Energy and Throughput for Caches in MPSoCs

Arsalan Shahid[1], Muhammad Tayyab[2], Muhammad Yasir Qadri[3], Nadia N. Qadri[4], and Jameel Ahmed[2]

1. School of Computer Science, University College Dublin, Belfield, Dublin 4, Ireland
2. Department of Electrical Engineering, HITEC University, Taxila Cantt., Pakistan
3. University of Essex, Colchester CO4 3SQ, UK
4. COMSATS Institute of Information and Technology, Wah Cantt., Pakistan

[arsalan.shahid@ucdconnect.ie, muhammad.tayyab@hitecuni.edu.pk, yasirqadri@acm.org, drnadia@ciitwah.edu.pk, jameel@hitecuni.edu.pk]

**ABSTRACT**

General trends in computer architecture are shifting more towards parallelism. Multicore architectures have proven to be a major step in processors evolution. With the advancement in Multicore architecture researchers are focusing finding different solutions to fully utilize the power of multiple cores. With ever increasing number of cores on a chip, the role of cache memory has become pivotal. An ideal memory configuration should be both large and fast, however in fact system architects have to strike a balance between the size and access time of the memory hierarchy. It is important to know the impact of a particular cache configuration on the throughput and energy consumption of the system at design time. This paper presents an enhanced version of previously proposed cache energy and throughput models for multicore systems. These models use significantly a smaller number of input parameters as compared to other models. This paper also validates the proposed models through cycle accurate simulator and a renowned processor power estimator. The results show that the proposed energy models provide an accuracy within a maximum error range of 10% for single core processors and around 5% for MPSoCs, and the throughput models result in maximum error of up to 11.5% for both single and multicore architectures.

*Keywords:*
*Cache mathematical models, energy and throughput models, multilevel cache, multicore*

## 1. INTRODUCTION

Cache memories are an integral part of most modern processor architectures. For a processor architect choice of components such as cache size and associativity, pipeline depth, number of cores, instruction set design is a critical decision to make. In most cases verification methodology based on Transaction Level Modeling (TLM) [4] or virtualized platforms [16] is used to analyze a proposed configuration several times. However, in general these tools and methodologies are unable to evaluate power consumption of a particular configuration. A single configuration can take several hours for complete evaluation. This process is called design space exploration and is often considered a design time, offline technique. On the other hand, in case of energy aware reconfigurable architectures, an early decision is often required to evaluate impact of a particular configuration beforehand [12] [23]. In this case lightweight analytical models are often required in order to assist the reconfiguration engine.

In either case; i.e. for throughput and energy aware hardware exploration at design time or reconfiguration at run-time it is imperative to gauge the performance of cache architectures so as to evaluate their impact on energy requirement and throughput of the system. This paper presents multicore extension of previously proposed cache energy and throughput models [17] [21] [22] [23]. These models require fewer inputs, obtainable using functional simulations and provide an accurate estimate of timings and energy consumption of the cache architecture. The proposed models analyze the energy and throughput of multicore cache hierarchies per application basis thus providing the hardware and software designer with the feedback vital to tune the cache or application for a given energy budget at both offline or online. This paper extends the state of the art by the following contributions:

 1. Multicore extension of the previously proposed models with the integration of Cycles per instructions (CPI) of a system.
 2. The models are evaluated over state-of-the-art Intel XEON series processors.
 3. The models have been validated by using HP Labs' McPAT (Multicore Power, Area and timing modeling Framework) [14] and MARSS-x86 (Micro Architectural and system simulator for Multicore Processors) [18] Simulators.

The aim of this research is to propose and validate the simplified mathematical models for energy and throughput of multicore, multilevel caches for application in the proposed multicore reconfigurable architecture [24].

The rest of this paper is divided into five sections. In the following section related work is discussed. The energy and throughput models for multicore cache are introduced in section 3. In section 4 the models are validated using a two-level cache hierarchy in multicore architecture, and the final section presents the conclusion.

## 2. RELATED WORK

This section presents the related research in the area of cache performance estimation, its usage for various applications and tools such as full system simulators and virtual platforms.

Basmadjian et al. [2] presented a methodology for estimating the power consumption of multicore processors by the resource sharing and power saving mechanisms. The authors propose component-based power models for multicore processors but used fixed capacitance model for the different components of processors and their approach was not extended for processors consisting of more than four cores. Lee at al. [13] have proposed a performance and power estimation technique (PET) for multicore systems. The scheme is based on accurate performance and power transformation model which predicts the performance and power consumption. Furthermore, it also gives the runtime configuration of multi-threaded applications. The results were compiled on an Intel Q6600 quad-core processor under two different frequency levels. The average estimated error of 2.1%-8.3% and 3.2%-6.5% over the measured data, respectively. Their work was limited to predict power consumption processor and do not determine energy of each component. Kamble et al. in [28] also presented detailed cache energy model. The analytic models for conventional caches were found to be accurate to within 2% error. However, their technique over predicts the power dissipation of low power caches by as much as 30%.

Dev at al. [4] devised post-silicon power mapping and modeling of multicore processors by using infrared imaging and performance counter measurements. An accurate finite element model that relates power consumption to temperature has been devised, along with compensating for the artifacts brought together by using infrared-transparent heat removal techniques. A standard numerical technique has been proposed to accurately translate thermal maps for heat sink system. Also, the designers formulated precise empirical models that estimate the infrared-based per-block power maps by means of the PMC measurements. These PMC models exactly estimate the transient power consumption of different processor blocks for that the SPEC CPU2006 benchmarks has been used. Kasichayanula et al. [11] proposed an idea of identifying power consumption accurately by developing Activity-based Model for GPUs (AMG). The core idea handled here is real-time power consumption, which is done by accurately estimated by using NVIDIAs Management Library (NVML). Model validation is done using Kill-A Watt power meter. The authors have claimed that the results are accurate within 10%. The models presented in their work holistically analyze the embedded system power and do not estimate energy consumption for individual components of a processor.

Pricopi et al. [20] proposed a software-based modeling technique for multiple types of cores, which can accomplish performance estimation and power consumption of workloads. The evaluation of the estimation framework technique was done on real asymmetric multicore ARM big.LITTLE asymmetric multi-core platform [6]. The model predicts the power performance behavior of an application on a target core, given all the specifications on run time. Whereas, the cores share the similar ISA but have heterogeneous micro-architecture. However, the work does not address scalability for multi-threaded applications.

Lim et al. [15] proposed a set of equations to estimate accurately worst-case time analysis (WCTA) for RISC processors. Their models include the details of the pipelining, instruction cache and data cache effects on real timeliness of the system. But the size of the program for analysis is still limited.

Taha et al. [25] presented an instruction throughput model of Superscalar processors. Their model includes parameters such as superscalar width, depth of pipeline, instruction fetch mechanism (in-order/out-of-order), branch predictor, central issue window width, number of functional units their latencies and throughputs, re-order buffer width and cache size and latency etc. Their model resulted in errors up to 5.5% when compared to the Simple Scalar simulator [1]. Wada et al. [26] proposed detailed circuit level analytical access time model for on-chip cache memories. The model takes inputs such as number of tag/data array per word/bit line etc. On comparing with SPICE results the model gives 20% error for an 8ns access time cache memory.

Yourst [27] developed PTLsim (A full system clock accurate simulator) to simulate each component at instruction level. This simulator features the configurable RTL level architecture and pipelines at the speed of host system. MARSS-x86 [18] is a cycle accurate complete system simulator for x86 and x64 based architectures, especially for multi-core hardwares. MARSS-x86 extends the functionality and support of PTLsim including complete user space simulations, unmodified software and OS stack and unmodified kernel. For Power consumption estimation for complete system, a tool named McPAT is developed by Li et al. [14] at HP Labs. This tool supports power estimation for various architectures including caches, NOC, multiprocessor, in-order, out-of-order, shared caches and integrated memory. The power consumption estimation is done at circuit level hence it is closer to the real system.

The following section presents the proposed cache energy and throughput models that can be used to get an accurate energy consumption and throughput estimates of a multicore architecture.

## 3. THE CACHE ENERGY AND THROUGHPUT MODELS

This section presents the energy and throughput models for a two-level cache hierarchy for multicore architectures.

### 3.1 Energy Models

If $E_{ic}$, $E_{dc}$, and $E_{l2c}$ is the energy consumed by instruction, data and level 2 (L2) cache operations, $E_{misc}$ is the Energy consumed by the instructions which do not require data memory access, CPI is the number of cycles per instruction and $E_{leak}$ the leakage energy of the processor. In the previous work CPI was considered to be 1, however in real-time scenarios it could vary depending on various parameters such as branching, predictions, parallelism and no. of cores per chip. CPI directly affects the energy consumption as shown in model below. The total energy consumption of the code $E_{total}$ in Joules [J] can be defined as,

$$E_{total} = (E_{ic} + E_{dc} + E_{l2c} + E_{misc.} + E_{leak})/CPI$$

Where,
L1 Instruction Cache
$$E_{ic} = E_{ic-read} + E_{ic-mp}$$
$$E_{ic-read} = E_{ic-rcycle} \cdot \eta_{ic-read}$$
$$E_{ic-mp} = E_{cycle} \cdot P_{ic-rmiss} \cdot \eta_{ic-rmiss}$$

L1 Data Cache
$$E_{dc} = E_{dc-read} + E_{dc-write} + E_{dc-mp}$$
$$E_{dc-read} = E_{dc-rcycle} \cdot \eta_{dc-read}$$
$$E_{dc-write} = E_{dc-wcycle} \cdot \eta_{dc-write}$$

$$E_{dc-mp} = E_{cycle} \cdot (P_{dc-rmiss} \cdot \eta_{dc-rmiss} + P_{dc-wmiss} \cdot \eta_{dc-wmiss})$$

L2 Cache
$$E_{l2c} = E_{l2c-read} + E_{l2c-write} + E_{dc-mp} + E_{l2c \to ram} + E_{l2c \to rom}$$

$$E_{l2c-read} = E_{l2c-rcycle} \cdot (\eta_{l2c-if} + \eta_{l2c-dread})$$
$$E_{l2c-write} = E_{l2c-wcycle} \cdot \eta_{l2c-dwrite}$$

$$E_{l2c-mp} = E_{cycle} \cdot \{P_{l2c-rmiss} \cdot (\eta_{l2c-if} + \eta_{l2c-dread}) + P_{l2c-wmiss} \cdot \eta_{l2c-dwrite}\}$$

In the above equations $E_{x-read}$, $E_{x-write}$, and $E_{x-mp}$ denote the read, write and miss penalty energy of the corresponding cache $x$ (i.e. instruction, data or L2 cache). The read and write cycle energy per cache access is denoted by $E_{x-rcycle}$ and $E_{x-wcycle}$. The number of data read and write transactions of the cache (including all hits and miss) is denoted by $\eta_{x-read}$ and $\eta_{x-write}$. Furthermore $\eta_{l2c-if}$, $\eta_{l2c-dread}$, $\eta_{l2c-dwrite}$ denote the L2 cache's instruction fetch, data read and data write transactions respectively. The processor's per cycle energy consumption is denoted by $E_{cycle}$, $P_{x-rmiss}$, $P_{x-wmiss}$, $\eta_{x-rmiss}$ and $\eta_{x-wmiss}$ denote the read/write miss penalty (in terms of number of cycles) and their corresponding miss rates. The energy consumed in L2 cache to data and code memory is denoted by $E_{l2c \to ram}$ and $E_{l2c \to rom}$ that could also be calculated by multiplying the number of memory accesses with their read and write cycles energy.

The idle mode leakage energy of the processor $E_{leak(std)}$ can be calculated as
$$E_{leak} = P_{leak} \cdot t_{idle}$$
Where $t_{idle}$ [Sec] is the total time for which processor was idle.

### 3.2 Throughput Models

If $t_{ic}$, $t_{dc}$, and $t_{l2c}$ is the time taken in instruction, data and level 2 (L2) cache operations, and $t_{ins}$ the time taken in execution of cache access instructions [Sec], $t_{x-read}$, $t_{x-write}$ and $t_{x-mp}$ the time taken in read, write and miss penalty for cache $x$; then $T_{total}$ the total time taken by an application could be estimated as
$$T_{total} = t_{ic} + t_{dc} + t_{l2c} + t_{ins}$$
Furthermore,
L1 Instruction Cache
$$t_{ic} = t_{ic-read} + t_{ic-mp}$$
$$t_{ic-mp} = t_{cycle} \cdot P_{ic-rmiss} \cdot \eta_{ic-rmiss}$$
$$t_{ic-mp} = t_{cycle} \cdot P_{ic-rmiss} \cdot \eta_{ic-rmiss}$$
L1 Data Cache
$$t_{dc} = t_{dc-read} + t_{dc-write} + t_{dc-mp}$$
$$t_{dc-read} = t_{dc-rcycle} \cdot \eta_{dc-read}$$
$$t_{dc-write} = t_{dc-wcycle} \cdot \eta_{dc-write}$$

$$t_{dc-mp} = t_{cycle} \cdot (P_{dc-rmiss} \cdot \eta_{dc-rmiss} + P_{dc-wmiss} \cdot \eta_{dc-wmiss})$$

L2 Cache
$$t_{l2c} = t_{l2c-read} + t_{l2c-write} + t_{dc-mp} + t_{l2c \to ram} + t_{l2c \to rom}$$
$$t_{l2c-read} = t_{l2c-rcycle} \cdot (\eta_{l2c-if} + \eta_{l2c-dread})$$
$$t_{l2c-write} = t_{l2c-wcycle} \cdot \eta_{l2c-dwrite}$$

$$t_{l2c-mp} = t_{cycle} \cdot \{P_{l2c-rmiss} \cdot (\eta_{l2c-if} + \eta_{l2c-dread}) + P_{l2c-wmiss} \cdot \eta_{l2c-dwrite}\}$$

And
$$t_{ins} = t_{cycle} \cdot \eta_{cycle} - t_{ic-read}$$

Where $t_{x-rcycle}$, $t_{x-wcycle}$ is the time taken per cache read and write cycle and $t_{cycle}$ is the processor cycle time in seconds [sec].

Table 1: Benchmark Applications from SPLASH-2

| Benchmark | Description |
|---|---|
| Barnes | Interaction of a system of bodies (galaxies or particles, for example) in three dimensions over a number of time steps |
| FMM | Simulates interactions in two dimensions using a different hierarchical N-body method called the adaptive Fast Multipole |
| Ocean | Studies large-scale ocean movements based on eddy and boundary currents |
| Water-Nsquared | Evaluated forces and potential that occur over time in a system of water molecules using predictor-corrector method |
| Water-Spatial | Imposes a 3-d spatial data structure on the cubical domain to solve molecular dynamics N-body problem |

## 4. MODEL VALIDATION

### 4.1 Simulation Setup

To validate the accuracy of the proposed models, MARSS-x86 [27] was used to run a number of benchmark applications from SPLASH-2 [29] bench-marking suite (see Table 1.) Three different type of Intel XEON Processors were used for evaluation purpose i.e. a Single Core XEON Foster [10], a dual core XEON E5503 [8], and a quad core XEON E5507 [14]. The parameters for each processor are mentioned in Table 2. The cache energy and throughput models discussed in section 3 require parameters such as $E_{x-rcycle}$, $E_{x-wcycle}$, $t_{x-rcycle}$, $t_{x-wcycle}$, that were obtained using HP labs' CACTI that is an integrated cache timing, power, and area model tool [3] (see Table 3.)

It is to be noted that MARSS-x86 provides cycle accurate simulation and timing information whereas for power estimation some external tool is required. HP labs has developed one such tool called, McPAT (Multicore Power, Area, and Timing) integrated power, area, and timing modeling framework for multithreaded, multicore, and manycore architectures was used for estimating Energy of various XEON processor models [14]. McPAT accepts simulation results from MARSS-x86 and then provides accurate power consumption estimates for a particular processor model.

Table 2: Processor Parameters

| Parameter | Value | | |
|---|---|---|---|
| | Single core | Dual core | Quad core |
| Brand | XEON | XEON | XEON |
| Model | Foster | E5503 | E5507 |
| Cores | 1 | 2 | 4 |
| Power [W] | 80 | 80 | 80 |
| Technology [nm] | 180 | 45 | 45 |
| L2 Cache [KBytes] | 256 | 256 | 256 |
| Clock Rate [MHz] | 2000 | 2000 | 2200 |

Table 3: CACTI (Cache Simulator) Data

| L1 I & D Cache | | L2 Cache | |
|---|---|---|---|
| Parameter | Value | Parameter | Value |
| Cache Size [KBytes] | 64 | Cache Size [KBytes] | 256 |
| Line Size [Bytes] | 64 | Line Size [Bytes] | 64 |
| Number Of Lines | 1024 | Number Of Lines | 1024 |
| R/W Ports | 1 | R/W Ports | 1 |
| Associativity | 4 | Associativity | 4 |
| Read Ports | 0 | Read Ports | 0 |
| Write Ports | 1 | Write Ports | 1 |
| Access Time [ns] | 0.894 | Access Time [ns] | 0.988 |
| Cycle Time [ns] | 0.32 | Cycle Time [ns] | 0.40 |
| Read Energy [nJ] | 0.049 | Read Energy [nJ] | 0.064 |
| Write Energy [nJ] [For D Cache] | 0.0081 | Write Energy [nJ] | 0.0137 |

### 4.2 Results

The energy model results for all the three XEON processor models are shown in Figure 1. The energy models for XEON Foster platform resulted in an error up to 10% in case of Ocean benchmark application whereas a minimal error of around 0.5% is observed in case of Barnes (see Figure 1 (b)). In case of multicore configurations maximum errors of up to 5% and 3% were observed for XEON E5503 and E5507 processors respectively (see Figure 1 (d,e)).

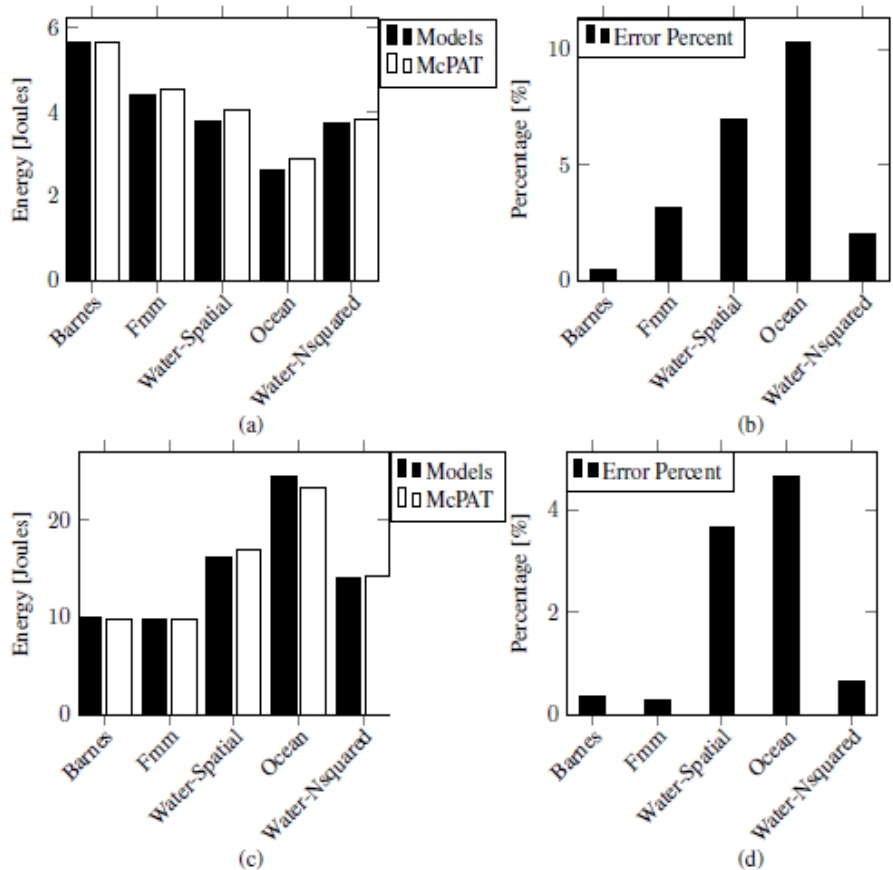

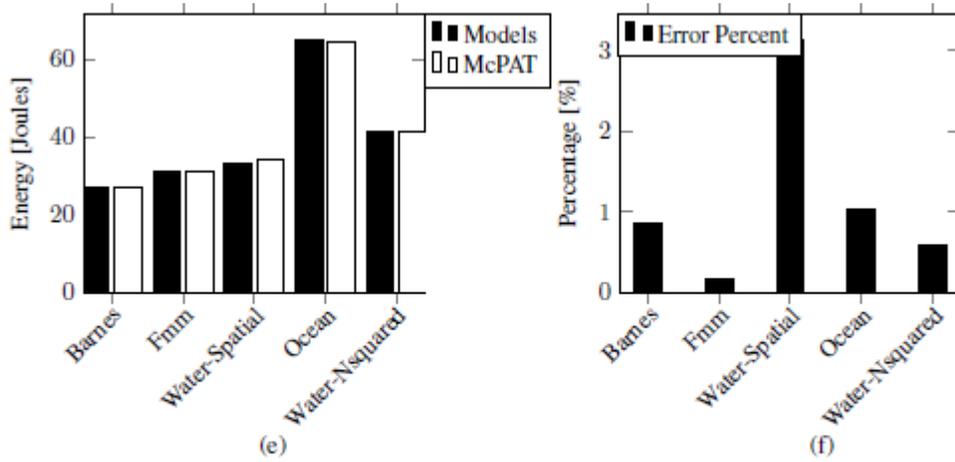

Fig. 1: Energy for various Benchmarks on SPLASH-2 Platform (a,b) Xeon Single Core (c,d) Xeon Dual Core (e,f) Xeon Quad Core

Figure 2 (a,b) show a comparative analysis of the throughput calculated from the presented throughput model (Predicted Throughput) and simulated throughput for XEON Foster Series (Single Core). Whereas Figure 2 (c,d) and Figure 2 (e,f) show the results for XEON E5503 (Dual Core) and XEON E5507 (Quad Core) respectively. The throughput models for Single Core XEON resulted in a maximum error up to 11.5% in case of FMM application, whereas a minimum error of around 3% is observed for Water-Spatial benchmark (see Figure 2 (b)). For the dual-core and quad-core models a maximum error of up to 8.5% and 11.5% is observed for Water-Spatial and Ocean applications respectively (see Figure 2 (d,f)).

It can be observed that the proposed models are able to estimate the energy and throughput of a multilevel cache hierarchy for both single core and multicore systems. The data obtained from CACTI [3] can be calculated by the same tool and stored in a look table, and the models can be used at runtime to estimate the effect of a cache on systems' throughput and performance. This scheme can be used for systems that support dynamic reconfiguration of memory system to make an early decision on cache sizing for a particular application in execution. One such example of the system is proposed by Qadri et. al. [24].

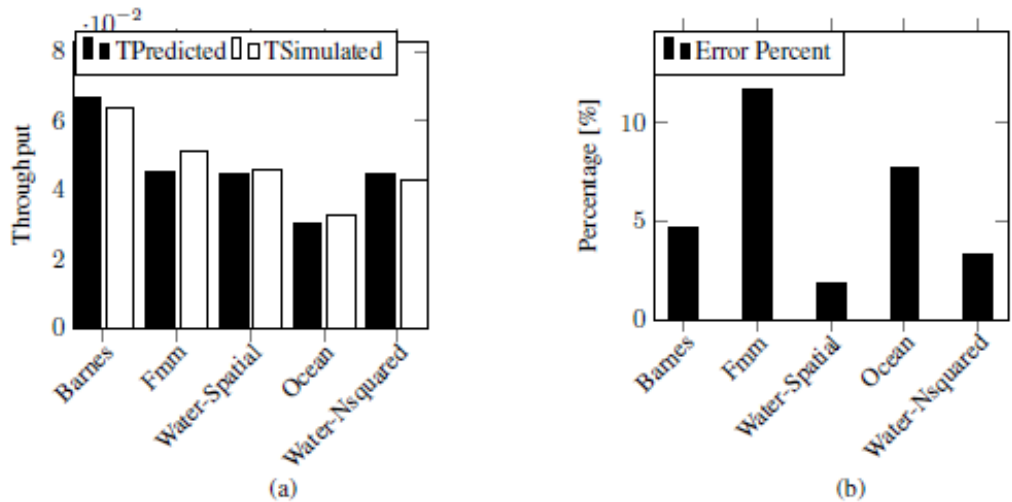

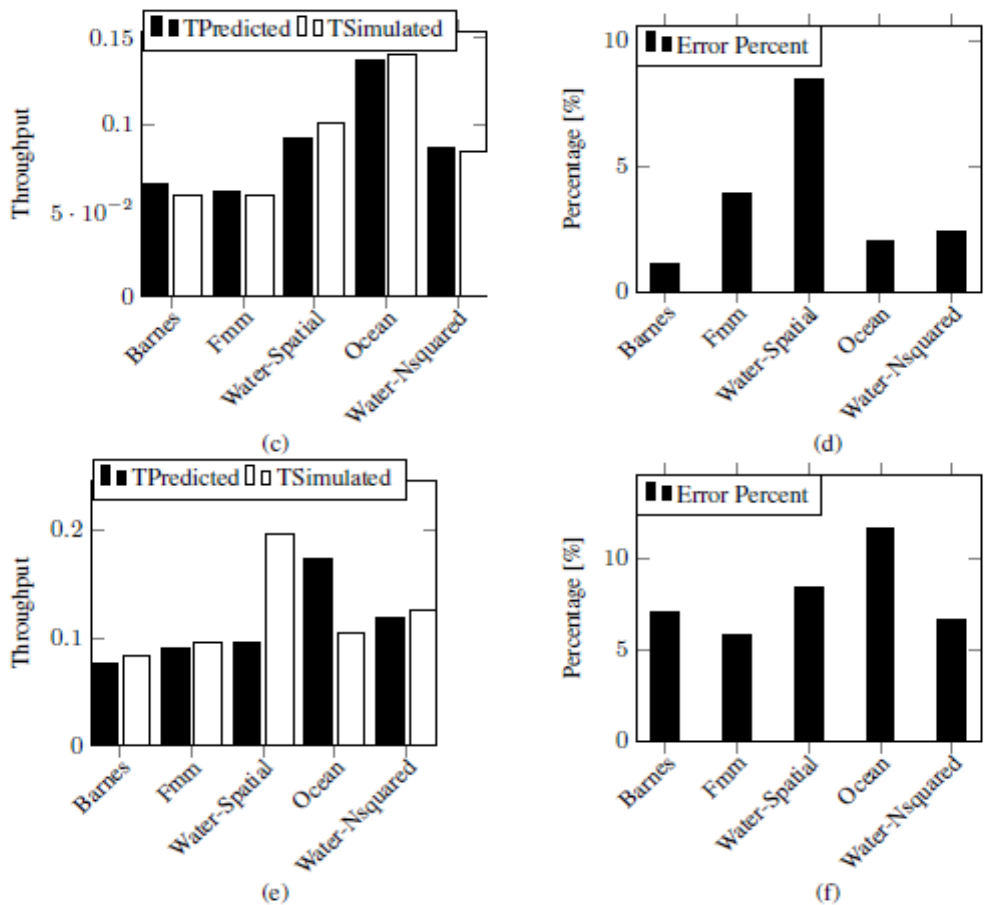

Fig. 2: Throughput for various Benchmarks on SPLASH-2 Platform (a,b) Xeon Single Core (c,d) Xeon Dual Core (e,f) Xeon Quad Core

## 5. CONCLUSION

In this paper multicore extension of previously presented cache energy and throughput models were presented. The models require a significantly smaller number of parameters as compared to the existing methods discussed in the related work. Moreover, these parameters can be easily obtained using the techniques adopted in the validation of the models. The models were validated with a two level cache model of XEON Foster, E5503 and E5507 processors, using standard benchmark applications and simulation tools. The cache energy models results were found to be only up to 10% deviated for XEON Foster, whereas for XEON E5503 and E5507 the error was 5% and 3% respectively; when compared with the simulators. Whereas for cache throughput models a maximum error of up to 11.5% is observed for both XEON Foster, and E5507. In the future work these models will be applied in real-time adaptive memory systems, where an accurate estimate of throughput and energy consumption for cache is required for reconfiguration purpose.

#### ACKNOWLEDGMENT

This work was supported by the National ICT R&D Fund, Pakistan through grant numbered: ICTRDF/TR&D/2012/65

### REFERENCES

1. Austin, T., Larson, E., Ernst, D.: *Simplescalar: An infrastructure for computer system modeling*. Computer 35(2), 59–67 (2002)
2. Basmadjian, R., de Meer, H.: *Evaluating and modeling power consumption of multi-core processors. In: Future Energy Systems: Where Energy, Computing and Communication Meet (e-Energy)*, 2012 Third International Conference on. pp. 1–10. IEEE (2012)
3. D. Tarjan, S.T., Jouppi, N.P.: *CACTI 4.0*. HP Laboratories Palo Alto (2006)
4. Dev, K., Nowroz, A.N., Reda, S.: *Power mapping and modeling of multi-core processors. In: Low Power Electronics and Design (ISLPED)*, 2013 IEEE International Symposium on. pp. 39–44. IEEE (2013)
5. Ferro, L., Pierre, L.: *ISIS: Runtime verification of TLM platforms. In: Advances in Design Methods from Modeling Languages for Embedded Systems and SoCs,* pp. 213–226. Springer (2010)
6. Greenhalgh., P.: *Big.little processing with arm cortex-a15 & cortex-a7: Improving energy efficiency in high-performance mobile platforms.* http://www.arm.com/files/downloads/big LITTLE Final Final.pdf,
7. Guthaus, M.R., Ringenberg, J.S., Ernst, D., Austin, T.M., Mudge, T., Brown, R.B.: *Mibench: A free, commercially representative embedded benchmark suite*. In: Workload Characterization, 2001. WWC-4. 2001 IEEE International Workshop on. pp. 3–14. IEEE (2001)
8. Intel Inc.: *Xeon e5503 datasheet*. available online at: http://ark.intel.com/products/37094/Intel-Xeon-Processor-E5503-4M-Cache-2_00-GHz-4_80-GTs-Intel-QPI
9. Intel Inc.: *Xeon e5507 datasheet*. available online at: http://ark.intel.com/products/37100/Intel-Xeon-Processor-E5507-4M-Cache-2_26-GHz-4_80-GTs-Intel-QPI
10. Intel Inc.: *Xeon foster datasheet*. available online at: http://ark.intel.com/products/codename/1923/Foster

11. Kasichayanula, K., Terpstra, D., Luszczek, P., Tomov, S., Moore, S., Peterson, G.D.: *Power aware computing on gpus.* In: Application Accelerators in High Performance Computing (SAAHPC), 2012 Symposium on. pp. 64–73. IEEE (2012)
12. Keramidas, G., Wong, S., Anjam, F., Brandon, A., Seedorf, et al.: *Embedded Reconfigurable Computing: the ERA Approach.* In: 11th IEEE International Conference on Industrial Informatics (2013)
13. Lee, Y.H., Kim, J.*: Fast and accurate on-line prediction of performance and power consumption in multicore-based systems.* In: Trust, Security and Privacy in Computing and Communications (TrustCom), 2013 12th IEEE International Conference on. pp. 1879–1886. IEEE (2013)
14. Li, S., Ahn, J.H., Strong, R.D., Brockman, J.B., Tullsen, D.M., Jouppi, N.P.: *Mcpat: an integrated power, area, and timing modeling framework for multicore and manycore architectures.* In: Microarchitecture, 2009. MICRO-42. 42nd Annual IEEE/ACM International Symposium on. pp. 469–480. IEEE (2009)
15. Lim, S.S., Bae, Y.H., Jang, G.T., Rhee, B.D., Min, S.L., Park, C.Y., Shin, H., Park, K., Moon, S.M., Kim, C.S.*: An accurate worst case timing analysis for risc processors. Software Engineering*, IEEE Transactions on 21(7), 593–604 (1995)
16. Magnusson, P.S., Christensson, M., Eskilson, J., et al.: *Simics: A full system simulation platform.* IEEE Computer 35, 50–58 (2002)
17. Muhammad Yasir, Q., Klaus D, M.M., et al.: *Data cache-energy and throughput models: design exploration for embedded processors.* EURASIP journal on embedded systems 2009 (2010)
18. Patel, A., Afram, F., Ghose, K.: *Marss-x86: A qemu-based micro-architectural and systems simulator for x86 multicore processors.* In: 1st International Qemu Users Forum. pp. 29–30 (2011)
19. Peuto, B.L., Shustek, L.J.*: An instruction timing model of cpu performance.* In: ACM SIGARCH Computer Architecture News. vol. 5, pp. 165–178. ACM (1977)
20. Pricopi, M., Muthukaruppan, T.S., Venkataramani, V., Mitra, T., Vishin, S.: *Power-performance modeling on asymmetric multi-cores.* In: Compilers, Architecture and Synthesis for Embedded Systems (CASES), 2013 International Conference on. pp. 1–10. IEEE (2013)
21. Qadri, M.Y., Maier, K.: *Data cache-energy and throughput models: a design exploration for overhead analysis. In: Proceedings of the Conference on Design and Architectures for Signal and Image Processing* (DASIP08). Citeseer (2008)
22. Qadri, M.Y., Maier, K.: *Towards increased power efficiency in low end embedded processors: can cache help?* In: Proceedings of the 4th UK Embedded Forum. Citeseer (2008)
23. Qadri, M.Y., McDonald-Maier, K.D.*: Analytical evaluation of energy and throughput for multilevel caches.* In: Computer Modelling and Simulation (UKSim), 2010 12th International Conference on. pp. 598–603. IEEE (2010)
24. Qadri, M.Y., McDonald Maier, K.D., Qadri, N.N.: *Energy and throughput aware fuzzy logic based reconfiguration for MPSoCs.* Journal of Intelligent and Fuzzy Systems 26(1), 101–113 (2014)
25. Taha, T.M.,Wills, D.S.: *An instruction throughput model of superscalar processors.* Computers, IEEE Transactions on 57(3), 389–403 (2008)
26. Wada, T., Rajan, S., Przybylski, S.A.: *An analytical access time model for on-chip cache memories. Solid-State Circuits*, IEEE Journal of 27(8), 1147–1156 (1992)
27. Yourst, M.T.: *Ptlsim: A cycle accurate full system x86-64 microarchitectural simulator.* In: Performance Analysis of Systems & Software, 2007. ISPASS 2007. IEEE International Symposium on. pp. 23–34. IEEE (2007)
28. Kamble, M.B., Ghose, K.: *Modeling energy dissipation in low power caches.* International Symposium on Low Power Electronics and Design, pp. 143–148 (1998)
29. Hughes, C.J., Pai, V.S., Ranganathan, P., Adve, S.V.*: Rsim: simulating shared-memory multiprocessors with ilp processors.* Computer 35(2), 40–49 (2002)